\documentclass[sigplan,10pt]{acmart}
\usepackage{xspace}
\usepackage{xstring}
\usepackage{graphicx}
\usepackage{booktabs}
\usepackage{multirow}
\usepackage{enumitem}
\usepackage{amsmath}
\usepackage{subcaption}
\usepackage[algo2e,ruled,lined,boxed,commentsnumbered, noend, linesnumbered, procnumbered]{algorithm2e} %
\usepackage{cleveref}
\usepackage[binary-units]{siunitx}
\usepackage[abbreviations]{foreign}
\usepackage{flafter} %
\usepackage{soul}
\usepackage[font=footnotesize]{caption}
\usepackage{titlesec}
\titlespacing\section{0pt}{11pt plus 4pt minus 2pt}{0pt plus 2pt minus 2pt}
\titlespacing\subsection{0pt}{11pt plus 4pt minus 2pt}{0pt plus 2pt minus 2pt}
\acmConference[EuroMLSys '23]{3rd Workshop on Machine Learning and Systems}{May 8, 2023}{Rome, Italy}
\acmYear{2023}\copyrightyear{2023}
\setcopyright{acmlicensed}
\acmBooktitle{3rd Workshop on Machine Learning and Systems (EuroMLSys '23), May 8, 2023, Rome, Italy}
\acmPrice{15.00}
\acmISBN{979-8-4007-0084-2/23/05}
\acmDOI{10.1145/3578356.3592587}
\startPage{1}

\usepackage{listings}
\usepackage{balance}

\usepackage{xcolor}     %
\definecolor{dkgreen}{rgb}{0,0.5,0}
\definecolor{green}{rgb}{0,0.5,0}
\definecolor{gray}{rgb}{0.5,0.5,0.5}
\definecolor{mauve}{rgb}{0.58,0,0.52}
\definecolor{deepred}{rgb}{0.6,0,0}

\lstdefinestyle{code}{
  language=Python,
  showstringspaces=false,
  columns=flexible,
  basicstyle={\ttfamily\small},
  numbers=left,
  numberstyle=\scriptsize\color{black},
  keywordstyle=\ttfamily\color{mauve},
  numbersep=5pt,
  commentstyle=\color{dkgreen},
  stringstyle=\color{mauve},
  breaklines=true,
  breakatwhitespace=false,
  tabsize=2,
  xleftmargin=0.4cm,
 frame=bottomline, 
  framexleftmargin=1.5em,
  framesep=2pt,
  escapechar=|,
  emph={DLNode,training,dataset,sharing,graph,communication,Node},          %
  emphstyle=\color{deepred},    %
}
\newcommand{\sys}{\textsc{decentralizepy}\xspace}
\newcommand{\Sys}{\textsc{DecentralizePy}\xspace}

\newcommand{\FL}{\ac{FL}\xspace}
\newcommand{\DL}{\ac{DL}\xspace}

\newcommand{\celeba}{CelebA\xspace}

\newcommand{\cifar}{CIFAR-10\xspace}

\newcommand{\sgd}{{\xspace}\ac{SGD}\xspace}
\newcommand{\dpsgd}{{\xspace}\ac{D-PSGD}\xspace}

\newcommand{\niid}{\ac{non-IID}\xspace}

\graphicspath{ {figures/} }
\usepackage{tikz}
\usepackage{pgfplots}
\usepackage[eulergreek]{sansmath}

\pgfplotsset{compat=newest}
\usepgfplotslibrary{external,units,colorbrewer,groupplots,fillbetween}
\tikzsetexternalprefix{figures/}
\tikzset{external/mode=list and make}
\usetikzlibrary{patterns}

\makeatletter
\begingroup\endlinechar=-1\relax
\everyeof{\noexpand}%
\edef\x{\endgroup\def\noexpand\homepath{%
        \@@input|"kpsewhich --var-value=HOME" }}\x
\makeatother

\newcommand{\inputplot}[2]{%
	\includegraphics{main-figure#2.pdf}
}

\newcommand{\newgroupwidth}[2]%
{\expandafter\xdef\csname groupwidth#1\endcsname{#2}}

\newcounter{groupwidth}
\newsavebox{\groupwidthbox}
\makeatletter
{\edef\groupnumber{#1}%
	\stepcounter{groupwidth}%
	\@ifundefined{groupwidth\thegroupwidth}{\pgfmathsetlengthmacro{\mywidth}{\linewidth/\groupnumber}}%
	{\expandafter\let\expandafter\mywidth\csname groupwidth\thegroupwidth\endcsname}%
	\begin{lrbox}{\groupwidthbox}%
		\tikzset{/pgfplots/width={\mywidth}}%
		\ignorespaces}%
	{\end{lrbox}%
	\usebox\groupwidthbox
	\pgfmathsetlengthmacro{\mywidth}{\mywidth + (\linewidth - \wd\groupwidthbox)/\groupnumber}
	\immediate\write\@auxout{\string\newgroupwidth{\thegroupwidth}{\mywidth}}}
\makeatother
\usepackage{acronym}
\acrodef{DL}{decentralized learning}
\acrodef{ML}{machine learning}
\acrodef{D-PSGD}{decentralized parallel stochastic gradient descent}
\acrodef{FL}{federated learning}
\acrodef{SGD}{stochastic gradient descent}
\acrodef{IID}{independent and identically distributed}
\acrodef{non-IID}{non independent and identically distributed}
\acrodef{RMSE}{root mean square error}
\acrodef{RMW}{random model walk}
\acrodef{GL}{gossip learning}
\acrodef{DWT}{discrete wavelet transform}
\acrodef{LAN}{local area network}
\acrodef{WAN}{wide area network}
\newboolean{showcomments}
\setboolean{showcomments}{true}
\ifthenelse{\boolean{showcomments}}
{ \newcommand{\mynote}[3]{
		\fbox{\bfseries\sffamily\scriptsize#1}
		{\small$\blacktriangleright$\textsf{\emph{\color{#3}{#2}}}$\blacktriangleleft$}}
	\newcommand{\zzz}[1]{{\setlength{\fboxsep}{2pt}\fcolorbox{black}{yellow}{\textsf{\emph{#1}}}}\xspace}}
{ \newcommand{\mynote}[3]{}
	\newcommand{\zzz}[1]{}}

\begin{document}
\date{}

\title{Decentralized Learning Made Easy with DecentralizePy}

\author{Akash Dhasade}
\affiliation{\institution{EPFL}\country{}}
\orcid{0000-0003-4362-5548}
\author{Anne-Marie Kermarrec}
\orcid{0000-0001-8187-724X}
\affiliation{\institution{EPFL}\country{}}
\author{Rafael Pires}
\orcid{0000-0002-7826-1599}
\affiliation{\institution{EPFL}\country{}}
\author{Rishi Sharma}
\orcid{0000-0002-1928-1549}
\affiliation{\institution{EPFL}\country{}}
\authornote{Corresponding author: \texttt{first.last}@epfl.ch}
\author{Milos Vujasinovic}
\affiliation{\institution{EPFL}\country{}}
\orcid{0009-0006-5083-3045}

\begin{abstract}
\Ac{DL} has gained prominence for its potential benefits in terms of scalability, privacy, and fault tolerance.
It consists of many nodes that coordinate without a central server and exchange millions of parameters in the inherently iterative process of \ac{ML} training.
In addition, these nodes are connected in complex and potentially dynamic topologies.
Assessing the intricate dynamics of such networks is clearly not an easy task.
Often in literature, researchers resort to simulated environments that do not scale and fail to capture practical and crucial behaviors, including the ones associated to parallelism, data transfer, network delays, and wall-clock time.
In this paper, we propose \sys, a distributed framework for decentralized \ac{ML}, which allows for the emulation of large-scale learning networks in arbitrary topologies. We demonstrate the capabilities of \sys by deploying techniques such as sparsification and secure aggregation on top of several topologies, including dynamic networks with more than one thousand nodes.
\end{abstract} 
\begin{CCSXML}
	<ccs2012>
	<concept>
	<concept_id>10003033.10003034.10003038</concept_id>
	<concept_desc>Networks~Programming interfaces</concept_desc>
	<concept_significance>500</concept_significance>
	</concept>
	<concept>
	<concept_id>10010147.10010919.10010172</concept_id>
	<concept_desc>Computing methodologies~Distributed algorithms</concept_desc>
	<concept_significance>300</concept_significance>
	</concept>
	<concept>
	<concept_id>10010147.10010257.10010321</concept_id>
	<concept_desc>Computing methodologies~Machine learning algorithms</concept_desc>
	<concept_significance>300</concept_significance>
	</concept>
	<concept>
	<concept_id>10010520.10010521.10010537.10010540</concept_id>
	<concept_desc>Computer systems organization~Peer-to-peer architectures</concept_desc>
	<concept_significance>500</concept_significance>
	</concept>
	</ccs2012>
\end{CCSXML}

\ccsdesc[500]{Networks~Programming interfaces}
\ccsdesc[300]{Computing methodologies~Distributed algorithms}
\ccsdesc[300]{Computing methodologies~Machine learning algorithms}
\ccsdesc[500]{Computer systems organization~Peer-to-peer architectures}

\keywords{decentralized learning, middleware, machine learning, distributed systems, peer-to-peer, network topology} 
\maketitle

\acresetall
\section{Introduction}

There has been a shift in \ac{ML} training, which is now taking place at the location where data is generated, instead of the earlier method of first moving the data to be processed in data centers.
\Ac{FL}~\cite{mcmahan2017communication} and \DL~\cite{lian2017dpsgd} have become prominent collaborative training methods that prioritize privacy, by solely exchanging updates of the model being trained.
In \ac{FL}, training is orchestrated by a central server where the server broadcasts the global model to participating nodes and later aggregates the updates received back from them.
Conversely, in \DL, nodes are connected in a fully decentralized communication topology.
Rather than relying on a server, nodes in \DL train on their local datasets and share the resulting models with neighboring nodes.
The aggregated models obtained from these exchanges lead to a converged global model at the end of the \DL training process.
While \ac{FL} has already received more attention from academia and industry~\cite{yang2018applied,MLSYS2019_bd686fd6,caldas2018leaf,kairouz2019advances}, \DL is also recently gaining a lot of traction~\cite{lian2017dpsgd, pmlr-v119-koloskova20a,koloskova2019decentralized, koloskova2020decentralized}.

The popularity of \ac{FL} has been partially enabled by the availability of several frameworks for simulation or emulation~\cite{abadi2016tensorflow,ziller2021pysyft,he2020fedml,liu2021fate,beutel2020flower,lai:2022:fedscale,roth2022nvidia}, which allow researchers to quickly implement and assess novel \ac{FL} algorithms.
Support for \DL configurations, on the other hand, is rare and limited in these existing frameworks~\cite{he2020fedml}.
\ac{FL} frameworks typically offer abstractions for client-server interactions which are not suitable for \DL, since nodes in \DL communicate only with their immediate neighbors.
These limitations call for frameworks that provide abstractions involving peer-to-peer communication channels and topology manipulation for collaborative \ac{ML} training.

\DL research has often focused on exploring the impact of topology on learning performance~\cite{9996833,vogels2022beyond}, which has led to simulations using various static and dynamic topologies during training~\cite{pmlr-v119-koloskova20a}.
Another important area of \DL research targets to improve communication efficiency through compression, which reduces the number of model parameters exchanged between \DL nodes~\cite{strom2015scalable,alistarh2018sparseconvergence,lin2018deep}.
This includes techniques such as sparsification~\cite{alistarh2018sparseconvergence} and quantization~\cite{alistarh2017quantization}, which are either not readily available in \ac{FL} frameworks, or incompatible with \DL.
The same happens in secure aggregation~\cite{10.1145/3133956.3133982}, which imposes additional challenges when deployed in \DL training.
\sys offers these capabilities, empowering researchers to explore and innovate on \DL research without being constrained by the limitations of existing \ac{FL} frameworks.

Although we are not exhaustive in implementing all aspects of \DL systems, we provide, along with reference implementations, a modular and extensible framework that aims at easing the prototyping and deployment of such systems.
We demonstrate this flexibility with various topologies, sparsification, and secure aggregation techniques in \DL training.
Our results provide insights to encourage more research on these areas in order to improve the practicality of \DL.

\paragraph{Contributions}
\begin{itemize}
	\item We present \sys, a novel decentralized learning framework, which allows researchers and practitioners to experiment with the components of \DL systems and deploy them in real-world settings.
	\item We showcase how \sys can be leveraged to assess the effects of several topologies, dynamic networks, sparsification techniques, and secure aggregation in \DL systems.
	\item \sys is modular, easily extendable, and open-source~\cite{sacs:2022:decpy} under \textit{MIT License}.
\end{itemize}

The remainder of this paper is organized as follows: \Cref{sec:framework} describes the framework internals and \Cref{sec:eval} showcases its usage in several scenarios. We survey related work in \Cref{sec:related} and conclude in \Cref{sec:conclusion}.

\section{\Sys}
\label{sec:framework}

\begin{figure}
	\centering
	\includegraphics{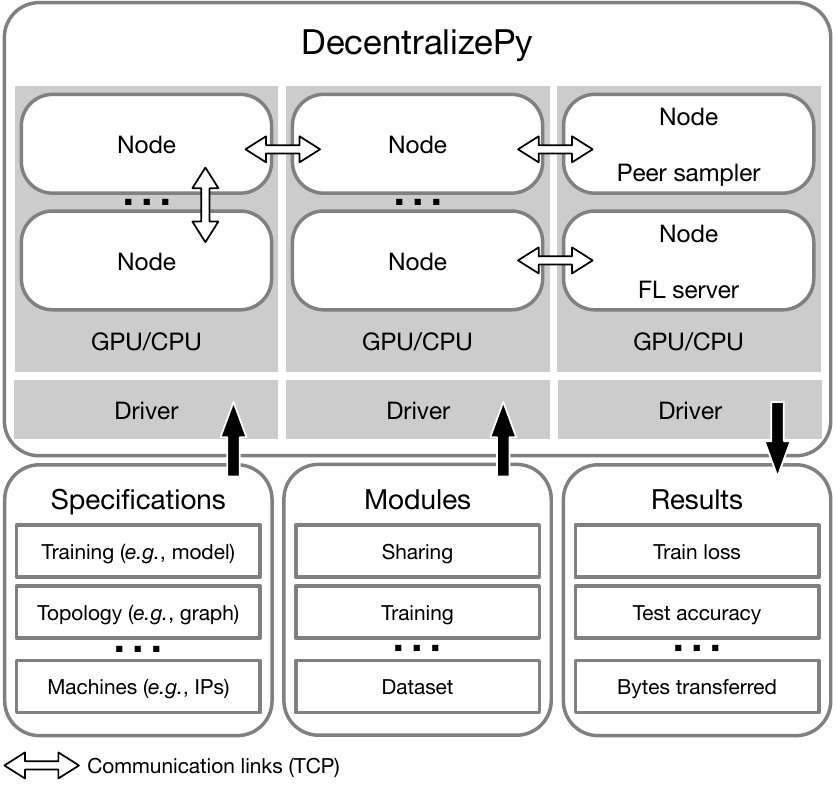}
	\caption{Overview of the \sys framework. Each node along with its driver may run on a separate machine. The driver takes as input the specifications and modules. The node dynamically loads the specified modules. Results are dumped locally and later aggregated. Nodes can be specialized to perform different tasks. In conventional \DL, we would have only basic \texttt{Node}(s). To emulate \FL, a node can be modified to coordinate the training, shown as the \FL server.}
	\label{fig:decpy_framework}
\end{figure}

\subsection{Design overview}
Works in \DL research %
are mostly evaluated in simulated environments.
These simulations are either done in scenarios where a single machine simulates all nodes (\ie, not scalable), or in cluster settings over MPI (\ie, constrained to the \ac{LAN})~\cite{9996833, vogels2022beyond, zhu2022topology, vogels2020powergossip}.
The design of \sys takes into account two key needs: the ability to quickly develop research prototypes, as well as actually deploying large-scale \DL systems.
We elaborate next on these design goals.

\paragraph{Modularity.}
\DL systems consist of multiple components, and conducting research in this field requires adjusting and testing them.
For generality, it is often necessary to evaluate different datasets and models when building learning systems.
In decentralized training, the overlay topology, \ie the way nodes are connected, is especially critical for achieving satisfactory model convergence~\cite{bars:2022:refined}.
Finally, the protocol itself, \ie who to communicate with, what is the message content, and how to aggregate the received parameters, varies among systems.
To facilitate the implementation of new \DL systems and experimentation with various ML aspects, including datasets, models, and topologies, \sys incorporates loosely coupled modules with an object-oriented design.
\Cref{sec:architecture} provides a more detailed explanation of these modules.

\paragraph{One-node one-process.}
In order to achieve rapid prototyping, researchers commonly favor simple implementations built from scratch.
In \DL systems, this often means simulations on a single machine or in a cluster environment~\cite{9996833, vogels2022beyond, zhu2022topology, vogels2020powergossip}.
Real-world deployment of such systems would however imply numerous nodes running on geographically-distant machines.
Unfortunately, \DL systems specifically designed for cluster environments are difficult to configure and scale beyond the \ac{LAN}.
\sys has a \textit{one-node as one-process} design principle, so that we are able to scale and measure realistic system metrics at the granularity of a node (or process).
In other words, one \DL node is represented as a single process.
These nodes communicate over network sockets and do not distinguish processes on the same or different machines.
This means that the same testbed can run in a cluster environment or on real-world machines over \acp{WAN} by just configuring the IP address information.
This approach simplifies the configuration of system parameters like the number of processes per node (according to, for instance, the number of CPU cores), network bandwidth, latency, and packet drop, enabling practitioners to study their systems in detail.
In addition, measurements of system metrics such as data transferred, memory usage, and CPU time become easier to collect on each node.
Furthermore, the emulation and deployment can reveal  behaviors that would not show up in simulations.
These benefits streamline the assessment of the performance and scalability of \DL systems.

\subsection{Architecture}
\label{sec:architecture}
In this section, we describe \sys modules and how they can be customized to develop new \DL systems.
\Cref{fig:decpy_framework} shows the architecture of the framework and how different modules are integrated.

\paragraph{Node.}
The \texttt{node} module acts as the skeleton code that performs the \DL task by instantiating and calling the methods of the rest of the modules.
An object of a sub-class of the \texttt{node} module is instantiated when a \DL process starts.
A \texttt{node} can be designed to perform a variety of tasks from being a \DL client to a \ac{FL} server or a centralized peer sampler.
The \texttt{node} module provides complete flexibility in creating and decoding messages to be exchanged with peers.
To get practitioners started with \sys, the framework is already equipped with the implementations of \DL clients, a centralized peer sampler, a parameter server, a \ac{FL} server, and secure aggregation clients as nodes.

\paragraph{Graph.}
The \texttt{graph} module manages the overlay network, which constrains the communication of nodes to only immediate neighbors.
This overlay graph can be modified at run time by the node, hence supporting both static and dynamic topologies.
The topology can be read from a graph file having edges or an adjacency list.
With this, we support swift switching of topologies, which could be generated by external libraries, during experimentation.
A graph file %
specifying the topology is shown as the \emph{topology specification} in \Cref{fig:decpy_framework}.

\paragraph{Model.}
This is a lightweight module inheriting the model class of the underlying ML framework.
The main purpose of having a \texttt{model} module is to store additional states.
For instance, this module allows the developer to store past gradients or how much the learning parameters changed in the last iteration.
This is especially convenient for some sparsification algorithms.
Since the models are quite specific to the datasets, these are implemented as part of the \texttt{dataset} modules.

\paragraph{Dataset.}
\ac{ML} frameworks must support a diverse range of datasets and models.
The \texttt{dataset} module of our framework provides this for a variety of learning tasks over multiple datasets and models, seamlessly integrating with the other modules.
Among the tasks it performs, we highlight: \emph{(i)}~reading the train and test sets; \emph{(ii)}~partitioning the datasets among nodes; \emph{(iii)}~evaluating the performance of a trained model on the test set; and \emph{(iv)}~specific model implementations.
\sys includes six datasets from the LEAF FL benchmark~\cite{caldas2018leaf} and \cifar~\cite{krizhevsky2014cifar} with both \ac{IID} and \ac{non-IID} data partitioning. %

\paragraph{Training.}
The \texttt{training} module performs local training steps of the model on the given training set.
The optimizer and loss function for the learning tasks are provided as part of the training specifications shown in \Cref{fig:decpy_framework}.
Having access to the instances of both \texttt{model} and \texttt{dataset} modules, the \texttt{training} module can modify the additional state variables in these modules for various purposes, \eg prioritizing certain weights during model compression.

\paragraph{Sharing and Communication}
In \DL, the nodes interact by exchanging messages.
The \texttt{sharing} module decides the contents of these messages and the aggregation procedure.
For model sharing, the messages would contain serialized parameters and the aggregation scheme will average the received models.
In the presence of sparsification (model compression), the messages would contain a subset of model parameters, and the aggregation scheme needs to account for missing parameters.
Implementations of sparsification schemes such as random sampling, \texttt{TopK}~\cite{alistarh2018sparseconvergence}, and \textsc{Choco-SGD}~\cite{koloskova2019decentralized} are available as \texttt{sharing} modules in \sys.
For data-sharing architectures~\cite{dhasade:2022:rex}, in turn, the \texttt{sharing} module would include raw data in the messages, and the aggregation procedure would append the received data to the local dataset.
These messages are passed on to \texttt{communication} by the \text{node} to send them to the correct recipients.
\sys also has an existing implementation of \texttt{communication} using ZeroMQ~\cite{boccassi:2022:zmq} over TCP.

\begin{figure}
\begin{lstlisting}
from decentralizepy.node.Node import Node

class DLNode(Node):
    def run(self, iterations, training, dataset, sharing, graph, communication):
        for round in range(iterations):
            training.train(dataset)
            msg = sharing.get_message()
            neighbors = graph.get_neighbors()
            communication.send(neighbors, msg)
            rcv = communication.receive_from_all()
            sharing.average(rcv)
            dataset.test()
\end{lstlisting}
\caption{A Python code snippet to demonstrate a simple \DL node using the modules of \sys colored in red. The node repeatedly trains its model on the local dataset (line 6), exchanges the model with the neighbors (lines 7-10), aggregates the models (line 11), and evaluates the average model on the test set (line 12).}
\label{fig:nodeCode}
\end{figure}

\begin{figure*}[t!]
	\centering
	\inputplot{plots/topologies_h}{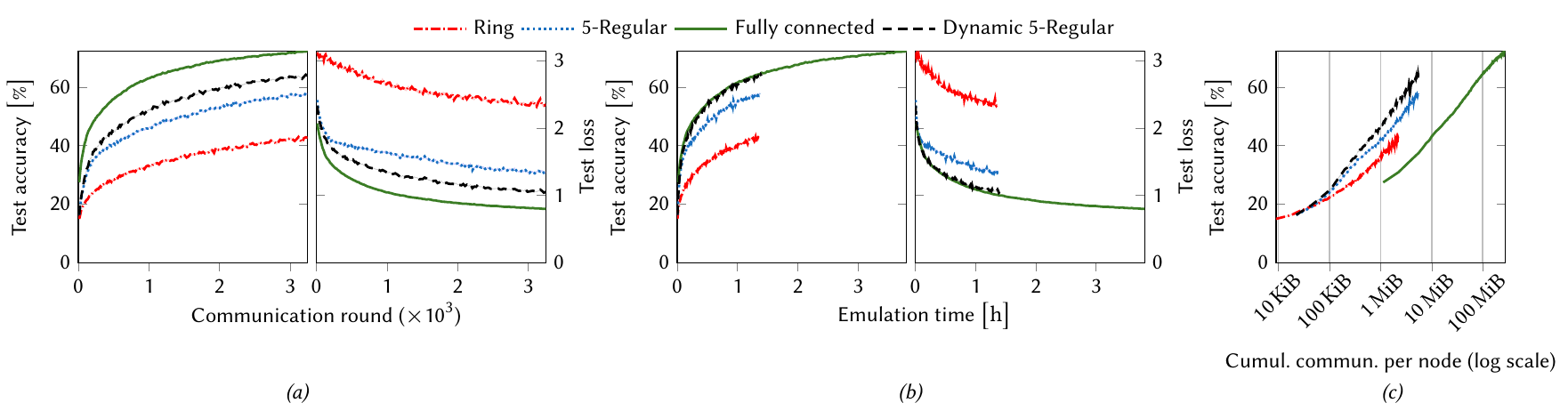}
	\caption{Performance of 256-node \DL across three topologies and a dynamic 5-regular graph. \emph{(a)} The denser the topology, the better the accuracy:  fully connected > d-regular > ring. They all run for the same number of communication rounds. \emph{(b)} When considering emulation time, fully connected takes the longest to perform the same number of rounds. \emph{(c)} Denser topologies incur significantly more communication costs. We observe that d-regular graphs offer a favorable tradeoff between accuracy, communication, and emulation time compared to \emph{ring} or \emph{fully connected}. Dynamic d-regular surprisingly matches the convergence of fully connected across time (b) at significantly lower communication cost (c).}
	\label{fig:topologies}
\end{figure*}

\paragraph{Mapping, Compression, and Utils}
There are some auxiliary modules to support the framework: (1) \texttt{Mapping}, (2) \texttt{compression}, and (3) \texttt{utils}.
To support both cluster environments and real-world deployment, \texttt{mapping} associates nodes with particular machines.
\texttt{Compression} module packages general-purpose compression algorithms for floating-point and integer lists.
\texttt{Utils} contains commonly used functions across the framework like Python dictionary manipulations and command-line argument parsing.

\sys allows the developer to instantiate a \texttt{node} and provide the implementations of the other modules separately, as shown in \Cref{fig:decpy_framework}.
Each module provides a base class that defines the interface for the module.
New implementations of modules can extend the base class and legacy implementations may be ported by wrapping them in the module interface.
These implementations are dynamically loaded to streamline the process of substituting the implementations for rapid prototyping.
Furthermore, since \sys is fully-decentralized, the nodes can have heterogenous datasets, different sharing strategies, optimizers, and learning rates for full flexibility.
\Cref{fig:nodeCode} shows how to write a simplified decentralized learning node using \sys.
The functions of the other modules (\eg \texttt{sharing}) invoked here can be overloaded to customize the system.
Given the decentralized nature of the nodes, each of them locally writes logs and results in JSON files.
To compute aggregate statistics of the system, we collect and process the results in a single machine at the end.

\subsection{Implementation}
Written in Python v3.8, the modules span a total of over \num{10000} lines of code.
Additionally, the framework inherits core \ac{ML} functionalities from PyTorch v1.10.2~\cite{torchSoftware}.
\sys contains several implementations of modules to facilitate building \DL systems as described earlier.
It is important to note that these are just reference implementations for a quick start in developing new \DL systems with the framework.
We call for the support of the scientific community to improve \sys by adding more implementations of modules derived from recent and upcoming research.

\section{Evaluation}
\label{sec:eval}
In this section, we demonstrate the power of \sys by implementing changing topologies, state-of-the-art sparsification algorithms, secure aggregation in \DL, and a scalability study.

\subsection{Experimental Setup}
The experiments are run on 16 hyperthreading-enabled machines equipped with 2 Intel(R) Xeon(R) CPU E5-2630 v3 @ 2.40GHz having 8 cores.
Based on the experiment, we run 48, 256, or 1024 \DL nodes (processes).
Each process is constrained to one logical CPU core in the experiments.
The nodes are oblivious about being on the same or different machines and communicate via TCP sockets using ZeroMQ~\cite{boccassi:2022:zmq}.
We tuned the learning-rate for the basic \sgd optimizer without momentum, and use it for training along with cross-entropy loss.
The \DL clients used \dpsgd~\cite{lian2017dpsgd} over Metropolis-Hastings weights~\cite{metropolis} for aggregation.
We use \cifar~\cite{krizhevsky2014cifar} dataset with a 2-sharding \ac{non-IID} data partitioning~\cite{mcmahan2017communication} which limits the number of classes per node to 4.
In addition to \cifar, we also use \celeba~\cite{caldas2018leaf} for secure-aggregation experiments.
The evaluation is done on the test-set of the respective datasets.
We run each experiment 5 times with different random seeds and present the average metrics with a 95\% confidence interval.

\subsection{Topologies and Dynamicity}

The performance of \DL is heavily influenced by the communication topology~\cite{koloskova2020decentralized}. 
We now demonstrate how \sys can flexibly simulate different topologies.
We start by running \sys with 256 learning nodes on a ring, d-regular with degree 5, and fully-connected static topologies.
The framework allows us to effortlessly change topologies for the experiments by only replacing the graph file described in \Cref{sec:architecture}.
To simulate dynamic topologies, \sys uses a centralized peer sampler which instantiates new topologies every round using the \texttt{graph} module.
Any dynamic graph can be realized within the peer sampler which then notifies each node of its neighbors.
We demonstrate a sample case where the peer sampler creates a random 5-regular topology every round.

\Cref{fig:topologies} shows the convergence plots with respect to the communication rounds, wall-clock time, and the cumulative bytes sent per node when run for the same number of communication rounds.
As one would expect, fully connected has the highest accuracy and the ring has the lowest accuracy at the end (\Cref{fig:topologies} (a)).
\sys also reveals that experiments with fully-connected topologies take nearly $3\times$ more time when compared to the rest for the same number of communication rounds (\Cref{fig:topologies} (b)).
This detail is often not evaluated in literature due to limited framework support.
We observe that d-regular topologies represent a favorable tradeoff between the extreme topologies (\Cref{fig:topologies} (b) and (c)).
Dynamic topologies perform much better than their static counterparts.
Moreover, dynamic 5-regular topology achieves almost identical accuracies to fully-connected given the same time deadline while having $51\times$ less communication cost.
Therefore, research should focus more on dynamic topologies to study their tradeoffs in more detail.

\begin{figure}[t]
	\centering
	\inputplot{plots/sparsification}{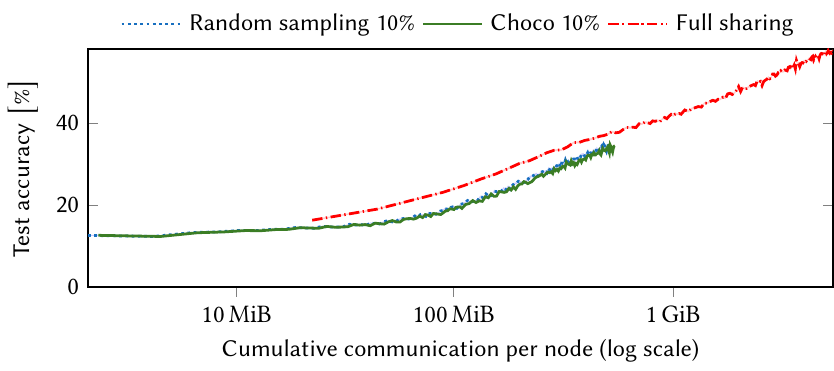}
	\caption{Performance of a 256-node \DL comparing the sparsification algorithms of \emph{random sampling} and \textsc{Choco-SGD} to \emph{full sharing}. The communication budget is set to 10\% and we run all algorithms for the same number of communication rounds. We observe that data non-IIDness and the scale of nodes significantly hurt the performance of sparsification algorithms. Under the same communication budget, full sharing tends to be robust and achieves higher accuracy.} 
	\label{fig:sparsification}
\end{figure}

\subsection{Sparsification}
\DL systems use sparsification algorithms to reduce the number of bytes exchanged in the network~\cite{alistarh2018sparseconvergence,lin2018deep}.
In sparsification, a communication budget specifies the percentage of model parameters shared by nodes with respect to full sharing, \ie sharing all parameters.
We now showcase how \sys supports commonly-used sparsification algorithms as part of the \texttt{sharing} module.
In full sharing, the basic \texttt{sharing} module generates a serialized parameter vector to be sent to neighboring nodes.
For sparsification, in contrast, we modify basic \texttt{sharing} to generate serialized tuples of the indices and values of the parameters chosen to be shared.
For the experiments, we consider a setup with a 5-regular topology of 256 learning nodes with a communication budget of $10\%$.

\Cref{fig:sparsification} shows the test accuracy vs. communication cost of two sparsification algorithms: (1) random sampling, and (2) hyperparameter-tuned state of the art \textsc{Choco-SGD}~\cite{koloskova2019decentralized} against the baseline of full sharing \DL.
The random sampling algorithm picks 10\% random parameters every round for sharing.
\textsc{Choco-SGD} uses sophisticated parameter-ranking and error-correction schemes to limit the loss of information due to sparsification.
We observe that sparsification loses too much information and performs significantly worse than full sharing.
This can be attributed to the complexity of the learning task. 
The convergence of sparsification algorithms drastically slows down in \ac{non-IID} settings with a large number of nodes.
To reach the same accuracy as both sparsification algorithms, full sharing uses significantly less communication.
Most works studying sparsification algorithms are often evaluated in IID settings with a limited number of nodes, resulting in overly optimistic estimations of the performance~\cite{hsiehskewscout2020}.
The use of \sys will enable researchers to study sparsification algorithms that are robust to difficult data distributions at scale.

\begin{figure}[t]
	\centering
	\inputplot{plots/fixed_budget_cesar_full_plot}{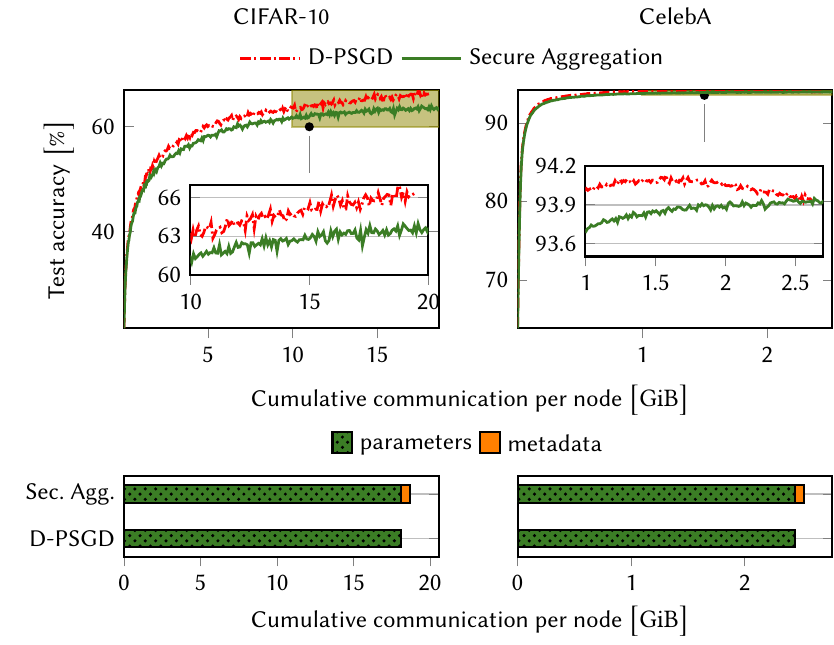}
	\caption{Performance of a 48-node \DL comparing Secure Aggregation with standard \DL without secure aggregation (\dpsgd). We observe that Secure Aggregation achieves comparable accuracy to \dpsgd on the CelebA dataset while it loses 3\% absolute accuracy on CIFAR-10 (first row). The privacy guarantees of Secure Aggregation come at a modest cost in additional communication (second row). } 
	\label{fig:sec_agg}
\end{figure}

\subsection{Secure Aggregation}
The well-established technique of secure aggregation~\cite{10.1145/3133956.3133982}, commonly used in centralized settings, ensures that participating nodes only have access to the aggregated model parameters, while keeping private the individual models of other training nodes.
This is achieved through the masking of parameters where pairs of nodes add cancellable masks to their respective models before sharing.
The receiver node upon aggregation gets the same aggregated model as one without secure aggregation. 
However, the masks prevent access to individual models.
Through a non-trivial implementation~\cite{vujasinovic2023secure}, we show that \sys can also be leveraged to perform such secure aggregation in the \DL setting.
We design the core procedure as part of the \texttt{node} module of \sys and conduct experiments with \num{48} nodes on \cifar and \celeba datasets for \num{10000} communication rounds.

\Cref{fig:sec_agg} shows the test accuracy vs. cumulative communication cost of \DL with and without secure aggregation.
Secure aggregation incurs approximately $3\%$ more communication due to metadata (shared seeds for pseudo-random number generation and masks) in addition to the parameters.
Furthermore, because masks and parameters are floating point numbers, there is a loss of precision which leads to $3\%$ loss in accuracy.

\begin{figure}[t]
	\centering
	\inputplot{plots/scalability}{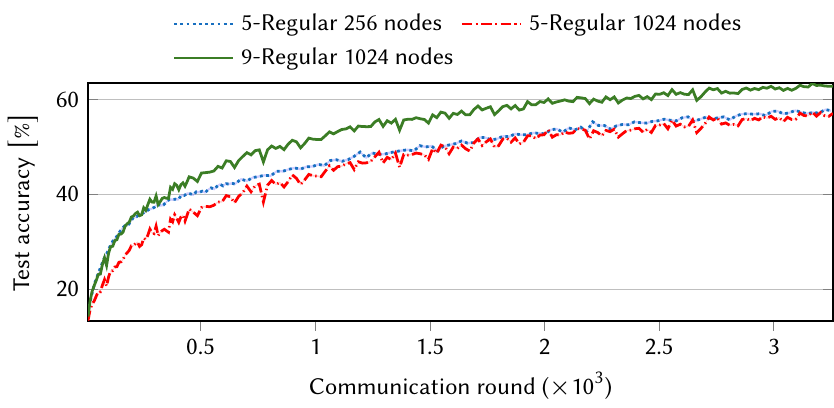}
	\caption{Performance of a 256-node and 1024-node \DL setups with varying degrees of d-regular topologies. Since the total dataset size remains the same when the nodes are scaled, each node receives $4\times$ fewer data samples in the 1024-node setup. We observe that 5-regular topology achieves the same performance in both 256- and 1024-node setups, suggesting that degree might be more influential than the number of data samples per node in \DL. Further increasing the degree to 9 benefits accuracy by almost 6\%.}
	\label{fig:scalability}
	
\end{figure}

\subsection{Scalability}
Finally, we perform a scalability study of \DL by increasing the number of nodes from 256 to 1024.
In this study, we particularly analyze the effect of the number of data samples per node and the degree of the node on the performance of \DL.
Since the underlying dataset size remains the same when the nodes are scaled, each node receives $4\times$ fewer data samples in the setup with 1024 nodes compared to the setup with 256 nodes.

We report the evolution of test accuracy for 256-node 5-regular, 1024-node 5-regular, and 1024-node 9-regular topologies in \Cref{fig:scalability}.
It is worth noting that the final performance of the 5-regular topology with \num{1024} and \num{256} nodes remain almost the same, even with \ac{non-IID} data and $4\times$ fewer data samples in the 1024-node setting.
Increasing the degree from $5$ to $9$ boosts the performance by $5.8$ accuracy points on average.
This suggests that the number of neighbors has a higher impact on learning than the number of data samples per node. 
Such novel insights are possible through advanced scalability studies, thanks to \sys. 
We highlight that \sys can run a much higher number of nodes relative to the relevant literature in decentralized learning systems~\cite{9996833, vogels2022beyond, zhu2022topology, vogels2020powergossip, koloskova2019decentralized}.

\section{Related work}
\label{sec:related}

Flower~\cite{beutel2020flower} is a framework for \ac{FL} emulation. Like \sys, Flower scales to one thousand nodes training in parallel. It is however limited to the \ac{FL} setup, \ie, it relies on a central server that orchestrates the \ac{ML} training by collecting and aggregating client models on every iteration. In addition, it does not offer seamless support for arbitrary communication topologies.

FedScale~\cite{lai:2022:fedscale} is both a collection of realistic datasets (with natural \ac{non-IID} and unbalanced data) and a runtime platform that supports cluster and mobile back-ends. In simulation mode, FedScale orchestrates execution requests over available resources (either single GPU/CPU or distributed) while keeping a \emph{client virtual clock}. This way, it achieves shorter evaluation times in comparison to the simulated scenario and reports up to \num{10000} training nodes in parallel on top of a cluster of 10 GPU nodes.  In addition, FedScale incorporates system speeds and availability traces of mobile devices to achieve realistic heterogeneity simulations. Unlike \sys, it has no support for \DL.

IPLS~\cite{pappas:2021:ipls} proposes \ac{ML} training that uses a decentralized filesystem as the communication channel. To reduce network traffic while ensuring some fault tolerance, they also propose model partitioning and replication of partitions across nodes. Each node only shares a particular set of layers (rather than the whole model) and each layer is potentially shared by several nodes. Its feasibility is yet to be confirmed, as authors show a very small deployment (50 nodes) with a simple learning task (MNIST) on top of \ac{IID} data. Moreover, IPLS depends on the underlying IPFS~\cite{benet:2014:ipfs} implementation and is therefore not flexible in terms of topology manipulation.

Kollaps~\cite{gouveia:2020:kollaps} is a decentralized network emulator that allows for the manipulation of end-to-end communication properties between nodes.
By shaping the network latency, bandwidth, packet loss, and jitter, they achieve very similar results of bare-metal deployments in geo-distributed setups. Kollaps is orthogonal and complementary to \sys, as it operates at the level of containers (\eg, Docker) and orchestrators (\eg, Kubernetes).

Closer to our proposal, FedML~\cite{he2020fedml} is a Federated Learning framework that provides several baseline implementations of models and datasets for reproducible research and benchmarking. FedML supports distinct topologies, including decentralized ones. Since it uses MPI~\cite{mpi40} for communication, it is primarily targeted to cluster environments. \sys, in contrast, can be smoothly deployed in \ac{WAN} and is more flexible than FedML in terms of dynamic topologies. 
\section{Conclusion}
\label{sec:conclusion}
Research in \DL systems requires efficient scaling to many nodes, ease of deployment, and a quick development cycle.
In this paper, we present \sys, a \DL framework designed to support these requirements.
Its modular design facilitates %
the replacement of components with customized ones, allowing rapid prototyping and deployment.

We demonstrate the power, flexibility, and scalability of \sys through a series of experiments across different topologies, sparsification algorithms, and privacy-preserving secure aggregation.
Using this framework, we were able to derive new insights concerning the effectiveness of dynamic topologies and the performance degradation of state-of-the-art sparsification algorithms on \niid learning tasks at scale.

\sys has been used by over a dozen students for projects in \DL.
Their feedback has been valuable in improving the usability and efficiency of the framework.
We expect the scientific community to also try and contribute to it. %
As future work, we plan to extend \sys with real-world traces of network traffic over Kollaps~\cite{gouveia:2020:kollaps}, client availability from FedScale~\cite{lai:2022:fedscale}, and decentralized peer sampling~\cite{jelasity2007gossip}.
We also plan to integrate additional state-of-the-art \DL algorithms, including quantization and topology construction.
Finally, using the framework as a platform, the community can build communication-efficient and privacy-preserving \DL systems that are robust to \niid data distributions at scale.

\bibliographystyle{references/ACM-Reference-Format}
\balance
\bibliography{references/main}


\begin{thebibliography}{39}

%%% the Web address.

\ifx \showCODEN    \undefined \def \showCODEN     #1{\unskip}     \fi
\ifx \showDOI      \undefined \def \showDOI       #1{#1}\fi
\ifx \showISBNx    \undefined \def \showISBNx     #1{\unskip}     \fi
\ifx \showISBNxiii \undefined \def \showISBNxiii  #1{\unskip}     \fi
\ifx \showISSN     \undefined \def \showISSN      #1{\unskip}     \fi
\ifx \showLCCN     \undefined \def \showLCCN      #1{\unskip}     \fi
\ifx \shownote     \undefined \def \shownote      #1{#1}          \fi
\ifx \showarticletitle \undefined \def \showarticletitle #1{#1}   \fi
\ifx \showURL      \undefined \def \showURL       {\relax}        \fi
% invisible to TeX
\providecommand\bibfield[2]{#2}
\providecommand\bibinfo[2]{#2}
\providecommand\natexlab[1]{#1}
\providecommand\showeprint[2][]{arXiv:#2}

\bibitem[Abadi et~al\mbox{.}(2016)]%
        {abadi2016tensorflow}
\bibfield{author}{\bibinfo{person}{Mart\'{\i}n Abadi}, \bibinfo{person}{Paul
  Barham}, \bibinfo{person}{Jianmin Chen}, \bibinfo{person}{Zhifeng Chen},
  \bibinfo{person}{Andy Davis}, \bibinfo{person}{Jeffrey Dean},
  \bibinfo{person}{Matthieu Devin}, \bibinfo{person}{Sanjay Ghemawat},
  \bibinfo{person}{Geoffrey Irving}, \bibinfo{person}{Michael Isard},
  \bibinfo{person}{Manjunath Kudlur}, \bibinfo{person}{Josh Levenberg},
  \bibinfo{person}{Rajat Monga}, \bibinfo{person}{Sherry Moore},
  \bibinfo{person}{Derek~G. Murray}, \bibinfo{person}{Benoit Steiner},
  \bibinfo{person}{Paul Tucker}, \bibinfo{person}{Vijay Vasudevan},
  \bibinfo{person}{Pete Warden}, \bibinfo{person}{Martin Wicke},
  \bibinfo{person}{Yuan Yu}, {and} \bibinfo{person}{Xiaoqiang Zheng}.}
  \bibinfo{year}{2016}\natexlab{}.
\newblock \showarticletitle{{TensorFlow}: a system for Large-Scale machine
  learning} \emph{(\bibinfo{series}{OSDI'16})}.
\newblock
\urldef\tempurl%
\url{https://www.usenix.org/conference/osdi16/technical-sessions/presentation/abadi}
\showURL{%
\tempurl}


\bibitem[Alistarh et~al\mbox{.}(2017)]%
        {alistarh2017quantization}
\bibfield{author}{\bibinfo{person}{Dan Alistarh}, \bibinfo{person}{Demjan
  Grubic}, \bibinfo{person}{Jerry~Z. Li}, \bibinfo{person}{Ryota Tomioka},
  {and} \bibinfo{person}{Milan Vojnovic}.} \bibinfo{year}{2017}\natexlab{}.
\newblock \showarticletitle{QSGD: Communication-Efficient SGD via Gradient
  Quantization and Encoding} \emph{(\bibinfo{series}{NIPS'17})}.
\newblock
\showISBNx{9781510860964}
\urldef\tempurl%
\url{https://proceedings.neurips.cc/paper_files/paper/2017/file/6c340f25839e6acdc73414517203f5f0-Paper.pdf}
\showURL{%
\tempurl}


\bibitem[Alistarh et~al\mbox{.}(2018)]%
        {alistarh2018sparseconvergence}
\bibfield{author}{\bibinfo{person}{Dan Alistarh}, \bibinfo{person}{Torsten
  Hoefler}, \bibinfo{person}{Mikael Johansson}, \bibinfo{person}{Sarit
  Khirirat}, \bibinfo{person}{Nikola Konstantinov}, {and}
  \bibinfo{person}{C\'{e}dric Renggli}.} \bibinfo{year}{2018}\natexlab{}.
\newblock \showarticletitle{The Convergence of Sparsified Gradient Methods}
  \emph{(\bibinfo{series}{NIPS'18})}.
\newblock
\urldef\tempurl%
\url{https://proceedings.neurips.cc/paper_files/paper/2018/file/314450613369e0ee72d0da7f6fee773c-Paper.pdf}
\showURL{%
\tempurl}


\bibitem[Bars et~al\mbox{.}(2023)]%
        {bars:2022:refined}
\bibfield{author}{\bibinfo{person}{Batiste~Le Bars},
  \bibinfo{person}{Aur{\'e}lien Bellet}, \bibinfo{person}{Marc Tommasi},
  \bibinfo{person}{Erick Lavoie}, {and} \bibinfo{person}{Anne-Marie
  Kermarrec}.} \bibinfo{year}{2023}\natexlab{}.
\newblock \showarticletitle{Refined Convergence and Topology Learning for
  Decentralized Optimization with Heterogeneous Data}
  \emph{(\bibinfo{series}{AISTATS'23})}.
\newblock
\showeprint[arXiv]{2204.04452}


\bibitem[Bellet et~al\mbox{.}(2022)]%
        {9996833}
\bibfield{author}{\bibinfo{person}{Aurélien Bellet},
  \bibinfo{person}{Anne-Marie Kermarrec}, {and} \bibinfo{person}{Erick
  Lavoie}.} \bibinfo{year}{2022}\natexlab{}.
\newblock \showarticletitle{D-Cliques: Compensating for Data Heterogeneity with
  Topology in Decentralized Federated Learning}
  \emph{(\bibinfo{series}{SRDS'22})}.
\newblock
\urldef\tempurl%
\url{https://doi.org/10.1109/SRDS55811.2022.00011}
\showDOI{\tempurl}


\bibitem[Benet(2014)]%
        {benet:2014:ipfs}
\bibfield{author}{\bibinfo{person}{Juan Benet}.}
  \bibinfo{year}{2014}\natexlab{}.
\newblock \showarticletitle{{IPFS} - Content Addressed, Versioned, {P2P} File
  System}.
\newblock \bibinfo{journal}{\emph{CoRR}} (\bibinfo{year}{2014}).
\newblock
\showeprint[arXiv]{1407.3561}


\bibitem[Beutel et~al\mbox{.}(2020)]%
        {beutel2020flower}
\bibfield{author}{\bibinfo{person}{Daniel~J Beutel}, \bibinfo{person}{Taner
  Topal}, \bibinfo{person}{Akhil Mathur}, \bibinfo{person}{Xinchi Qiu},
  \bibinfo{person}{Titouan Parcollet}, {and} \bibinfo{person}{Nicholas~D
  Lane}.} \bibinfo{year}{2020}\natexlab{}.
\newblock \showarticletitle{Flower: A Friendly Federated Learning Research
  Framework}.
\newblock  (\bibinfo{year}{2020}).
\newblock
\showeprint[arXiv]{2007.14390}


\bibitem[Boccassi et~al\mbox{.}(2023)]%
        {boccassi:2022:zmq}
\bibfield{author}{\bibinfo{person}{Luca Boccassi} {et~al\mbox{.}}}
  \bibinfo{year}{2023}\natexlab{}.
\newblock \bibinfo{title}{ZeroMQ: An open-source universal messaging library}.
\newblock
\newblock
\urldef\tempurl%
\url{https://zeromq.org}
\showURL{%
\tempurl}


\bibitem[Bonawitz et~al\mbox{.}(2019)]%
        {MLSYS2019_bd686fd6}
\bibfield{author}{\bibinfo{person}{Keith Bonawitz}, \bibinfo{person}{Hubert
  Eichner}, \bibinfo{person}{Wolfgang Grieskamp}, {et~al\mbox{.}}}
  \bibinfo{year}{2019}\natexlab{}.
\newblock \showarticletitle{Towards Federated Learning at Scale: System Design}
  \emph{(\bibinfo{series}{MLSys'19})}.
\newblock
\urldef\tempurl%
\url{https://proceedings.mlsys.org/paper_files/paper/2019/file/bd686fd640be98efaae0091fa301e613-Paper.pdf}
\showURL{%
\tempurl}


\bibitem[Bonawitz et~al\mbox{.}(2017)]%
        {10.1145/3133956.3133982}
\bibfield{author}{\bibinfo{person}{Keith Bonawitz}, \bibinfo{person}{Vladimir
  Ivanov}, \bibinfo{person}{Ben Kreuter}, \bibinfo{person}{Antonio Marcedone},
  \bibinfo{person}{H.~Brendan McMahan}, \bibinfo{person}{Sarvar Patel},
  \bibinfo{person}{Daniel Ramage}, \bibinfo{person}{Aaron Segal}, {and}
  \bibinfo{person}{Karn Seth}.} \bibinfo{year}{2017}\natexlab{}.
\newblock \showarticletitle{Practical Secure Aggregation for Privacy-Preserving
  Machine Learning} \emph{(\bibinfo{series}{CCS '17})}.
\newblock
\urldef\tempurl%
\url{https://doi.org/10.1145/3133956.3133982}
\showDOI{\tempurl}


\bibitem[Caldas et~al\mbox{.}(2019)]%
        {caldas2018leaf}
\bibfield{author}{\bibinfo{person}{Sebastian Caldas}, \bibinfo{person}{Peter
  Wu}, \bibinfo{person}{Tian Li}, \bibinfo{person}{Jakub Kone{\v{c}}n{\'y}},
  \bibinfo{person}{H.~Brendan McMahan}, \bibinfo{person}{Virginia Smith}, {and}
  \bibinfo{person}{Ameet Talwalkar}.} \bibinfo{year}{2019}\natexlab{}.
\newblock \showarticletitle{Leaf: A benchmark for federated settings}. In
  \bibinfo{booktitle}{\emph{2nd Intl. Workshop on Federated Learning for Data
  Privacy and Confidentiality}} \emph{(\bibinfo{series}{FL-NeurIPS'19})}.
\newblock
\showeprint[arXiv]{1812.01097}


\bibitem[Dhasade et~al\mbox{.}(2022)]%
        {dhasade:2022:rex}
\bibfield{author}{\bibinfo{person}{Akash Dhasade}, \bibinfo{person}{Nevena
  Dresevic}, \bibinfo{person}{Anne-Marie Kermarrec}, {and}
  \bibinfo{person}{Rafael Pires}.} \bibinfo{year}{2022}\natexlab{}.
\newblock \showarticletitle{{TEE}-based decentralized recommender systems: The
  raw data sharing redemption} \emph{(\bibinfo{series}{{IPDPS}'22})}.
\newblock
\urldef\tempurl%
\url{https://doi.org/10.1109/IPDPS53621.2022.00050}
\showDOI{\tempurl}


\bibitem[Gouveia et~al\mbox{.}(2020)]%
        {gouveia:2020:kollaps}
\bibfield{author}{\bibinfo{person}{Paulo Gouveia}, \bibinfo{person}{Jo\~{a}o
  Neves}, \bibinfo{person}{Carlos Segarra}, \bibinfo{person}{Luca Liechti},
  \bibinfo{person}{Shady Issa}, \bibinfo{person}{Valerio Schiavoni}, {and}
  \bibinfo{person}{Miguel Matos}.} \bibinfo{year}{2020}\natexlab{}.
\newblock \showarticletitle{Kollaps: Decentralized and Dynamic Topology
  Emulation} \emph{(\bibinfo{series}{EuroSys '20})}. Article
  \bibinfo{articleno}{23}.
\newblock
\showISBNx{9781450368827}
\urldef\tempurl%
\url{https://doi.org/10.1145/3342195.3387540}
\showDOI{\tempurl}


\bibitem[He et~al\mbox{.}(2020)]%
        {he2020fedml}
\bibfield{author}{\bibinfo{person}{Chaoyang He}, \bibinfo{person}{Songze Li},
  \bibinfo{person}{Jinhyun So}, \bibinfo{person}{Mi Zhang},
  \bibinfo{person}{Hongyi Wang}, \bibinfo{person}{Xiaoyang Wang},
  \bibinfo{person}{Praneeth Vepakomma}, \bibinfo{person}{Abhishek Singh},
  \bibinfo{person}{Hang Qiu}, \bibinfo{person}{Li Shen},
  \bibinfo{person}{Peilin Zhao}, \bibinfo{person}{Yan Kang},
  \bibinfo{person}{Yang Liu}, \bibinfo{person}{Ramesh Raskar},
  \bibinfo{person}{Qiang Yang}, \bibinfo{person}{Murali Annavaram}, {and}
  \bibinfo{person}{Salman Avestimehr}.} \bibinfo{year}{2020}\natexlab{}.
\newblock \showarticletitle{Fed{ML}: A research library and benchmark for
  federated machine learning}.
\newblock  (\bibinfo{year}{2020}).
\newblock
\showeprint{2007.13518}


\bibitem[Hsieh et~al\mbox{.}(2020)]%
        {hsiehskewscout2020}
\bibfield{author}{\bibinfo{person}{Kevin Hsieh}, \bibinfo{person}{Amar
  Phanishayee}, \bibinfo{person}{Onur Mutlu}, {and} \bibinfo{person}{Phillip~B.
  Gibbons}.} \bibinfo{year}{2020}\natexlab{}.
\newblock \showarticletitle{The Non-IID Data Quagmire of Decentralized Machine
  Learning}. In \bibinfo{booktitle}{\emph{Proceedings of the 37th International
  Conference on Machine Learning}} \emph{(\bibinfo{series}{ICML'20})}. Article
  \bibinfo{articleno}{408}.
\newblock
\urldef\tempurl%
\url{http://proceedings.mlr.press/v119/hsieh20a/hsieh20a.pdf}
\showURL{%
\tempurl}


\bibitem[Jelasity et~al\mbox{.}(2007)]%
        {jelasity2007gossip}
\bibfield{author}{\bibinfo{person}{M{\'a}rk Jelasity}, \bibinfo{person}{Spyros
  Voulgaris}, \bibinfo{person}{Rachid Guerraoui}, \bibinfo{person}{Anne-Marie
  Kermarrec}, {and} \bibinfo{person}{Maarten Van~Steen}.}
  \bibinfo{year}{2007}\natexlab{}.
\newblock \showarticletitle{Gossip-based peer sampling}.
\newblock \bibinfo{journal}{\emph{ACM Transactions on Computer Systems (TOCS)}}
  \bibinfo{volume}{25}, \bibinfo{number}{3} (\bibinfo{year}{2007}),
  \bibinfo{pages}{8--es}.
\newblock


\bibitem[Kairouz et~al\mbox{.}(2020)]%
        {kairouz2019advances}
\bibfield{author}{\bibinfo{person}{Peter Kairouz}, \bibinfo{person}{H.~Brendan
  McMahan}, \bibinfo{person}{Brendan Avent}, \bibinfo{person}{Aur\'{e}lien
  Bellet}, \bibinfo{person}{Mehdi Bennis}, \bibinfo{person}{Arjun
  Nitin~Bhagoji}, \bibinfo{person}{Kallista Bonawitz}, \bibinfo{person}{Zachary
  Charles}, \bibinfo{person}{Graham Cormode}, \bibinfo{person}{Rachel
  Cummings}, \bibinfo{person}{Rafael G.~L. D’Oliveira},
  \bibinfo{person}{Hubert Eichner}, \bibinfo{person}{Salim El~Rouayheb},
  \bibinfo{person}{David Evans}, \bibinfo{person}{Josh Gardner},
  \bibinfo{person}{Zachary Garrett}, \bibinfo{person}{Adri\`{a} Gasc\'{o}n},
  \bibinfo{person}{Badih Ghazi}, \bibinfo{person}{Phillip~B. Gibbons},
  \bibinfo{person}{Marco Gruteser}, \bibinfo{person}{Zaid Harchaoui},
  \bibinfo{person}{Chaoyang He}, \bibinfo{person}{Lie He},
  \bibinfo{person}{Zhouyuan Huo}, \bibinfo{person}{Ben Hutchinson},
  \bibinfo{person}{Justin Hsu}, \bibinfo{person}{Martin Jaggi},
  \bibinfo{person}{Tara Javidi}, \bibinfo{person}{Gauri Joshi},
  \bibinfo{person}{Mikhail Khodak}, \bibinfo{person}{Jakub Konecn\'{y}},
  \bibinfo{person}{Aleksandra Korolova}, \bibinfo{person}{Farinaz Koushanfar},
  \bibinfo{person}{Sanmi Koyejo}, \bibinfo{person}{Tancr\`{e}de Lepoint},
  \bibinfo{person}{Yang Liu}, \bibinfo{person}{Prateek Mittal},
  \bibinfo{person}{Mehryar Mohri}, \bibinfo{person}{Richard Nock},
  \bibinfo{person}{Ayfer \"{O}zg\"{u}r}, \bibinfo{person}{Rasmus Pagh},
  \bibinfo{person}{Hang Qi}, \bibinfo{person}{Daniel Ramage},
  \bibinfo{person}{Ramesh Raskar}, \bibinfo{person}{Mariana Raykova},
  \bibinfo{person}{Dawn Song}, \bibinfo{person}{Weikang Song},
  \bibinfo{person}{Sebastian~U. Stich}, \bibinfo{person}{Ziteng Sun},
  \bibinfo{person}{Ananda~Theertha Suresh}, \bibinfo{person}{Florian
  Tram\`{e}r}, \bibinfo{person}{Praneeth Vepakomma}, \bibinfo{person}{Jianyu
  Wang}, \bibinfo{person}{Li Xiong}, \bibinfo{person}{Zheng Xu},
  \bibinfo{person}{Qiang Yang}, \bibinfo{person}{Felix~X. Yu},
  \bibinfo{person}{Han Yu}, {and} \bibinfo{person}{Sen Zhao}.}
  \bibinfo{year}{2020}\natexlab{}.
\newblock \showarticletitle{Advances and open problems in federated learning}.
\newblock \bibinfo{journal}{\emph{Foundations and Trends in Machine Learning}}
  \bibinfo{volume}{14}, \bibinfo{number}{1--2} (\bibinfo{year}{2020}).
\newblock
\urldef\tempurl%
\url{https://doi.org/10.1561/2200000083}
\showDOI{\tempurl}


\bibitem[Koloskova et~al\mbox{.}(2020a)]%
        {koloskova2020decentralized}
\bibfield{author}{\bibinfo{person}{Anastasia Koloskova}, \bibinfo{person}{Tao
  Lin}, \bibinfo{person}{Sebastian~U Stich}, {and} \bibinfo{person}{Martin
  Jaggi}.} \bibinfo{year}{2020}\natexlab{a}.
\newblock \showarticletitle{Decentralized Deep Learning with Arbitrary
  Communication Compression} \emph{(\bibinfo{series}{ICLR'20})}.
\newblock
\urldef\tempurl%
\url{https://openreview.net/forum?id=SkgGCkrKvH}
\showURL{%
\tempurl}


\bibitem[Koloskova et~al\mbox{.}(2020b)]%
        {pmlr-v119-koloskova20a}
\bibfield{author}{\bibinfo{person}{Anastasia Koloskova},
  \bibinfo{person}{Nicolas Loizou}, \bibinfo{person}{Sadra Boreiri},
  \bibinfo{person}{Martin Jaggi}, {and} \bibinfo{person}{Sebastian Stich}.}
  \bibinfo{year}{2020}\natexlab{b}.
\newblock \showarticletitle{A Unified Theory of Decentralized {SGD} with
  Changing Topology and Local Updates} \emph{(\bibinfo{series}{ICML'20})}.
\newblock
\urldef\tempurl%
\url{https://proceedings.mlr.press/v119/koloskova20a.html}
\showURL{%
\tempurl}


\bibitem[Koloskova et~al\mbox{.}(2019)]%
        {koloskova2019decentralized}
\bibfield{author}{\bibinfo{person}{Anastasia Koloskova},
  \bibinfo{person}{Sebastian Stich}, {and} \bibinfo{person}{Martin Jaggi}.}
  \bibinfo{year}{2019}\natexlab{}.
\newblock \showarticletitle{Decentralized stochastic optimization and gossip
  algorithms with compressed communication} \emph{(\bibinfo{series}{ICML'19})}.
\newblock
\urldef\tempurl%
\url{https://proceedings.mlr.press/v97/koloskova19a.html}
\showURL{%
\tempurl}


\bibitem[Krizhevsky et~al\mbox{.}(2014)]%
        {krizhevsky2014cifar}
\bibfield{author}{\bibinfo{person}{Alex Krizhevsky}, \bibinfo{person}{Vinod
  Nair}, {and} \bibinfo{person}{Geoffrey Hinton}.}
  \bibinfo{year}{2014}\natexlab{}.
\newblock \showarticletitle{The CIFAR-10 dataset}.
\newblock  \bibinfo{volume}{55}, \bibinfo{number}{5} (\bibinfo{year}{2014}).
\newblock
\urldef\tempurl%
\url{https://www.cs.toronto.edu/~kriz/cifar.html}
\showURL{%
\tempurl}


\bibitem[Lai et~al\mbox{.}(2022)]%
        {lai:2022:fedscale}
\bibfield{author}{\bibinfo{person}{Fan Lai}, \bibinfo{person}{Yinwei Dai},
  \bibinfo{person}{Sanjay Singapuram}, {et~al\mbox{.}}}
  \bibinfo{year}{2022}\natexlab{}.
\newblock \showarticletitle{{F}ed{S}cale: Benchmarking Model and System
  Performance of Federated Learning at Scale}
  \emph{(\bibinfo{series}{ICML'22})}.
\newblock
\urldef\tempurl%
\url{https://proceedings.mlr.press/v162/lai22a.html}
\showURL{%
\tempurl}


\bibitem[Lian et~al\mbox{.}(2017)]%
        {lian2017dpsgd}
\bibfield{author}{\bibinfo{person}{Xiangru Lian}, \bibinfo{person}{Ce Zhang},
  \bibinfo{person}{Huan Zhang}, \bibinfo{person}{Cho-Jui Hsieh},
  \bibinfo{person}{Wei Zhang}, {and} \bibinfo{person}{Ji Liu}.}
  \bibinfo{year}{2017}\natexlab{}.
\newblock \showarticletitle{Can Decentralized Algorithms Outperform Centralized
  Algorithms? A Case Study for Decentralized Parallel Stochastic Gradient
  Descent} \emph{(\bibinfo{series}{NIPS'17})}.
\newblock
\urldef\tempurl%
\url{https://proceedings.neurips.cc/paper_files/paper/2017/file/f75526659f31040afeb61cb7133e4e6d-Paper.pdf}
\showURL{%
\tempurl}


\bibitem[Lin et~al\mbox{.}(2018)]%
        {lin2018deep}
\bibfield{author}{\bibinfo{person}{Yujun Lin}, \bibinfo{person}{Song Han},
  \bibinfo{person}{Huizi Mao}, \bibinfo{person}{Yu Wang}, {and}
  \bibinfo{person}{Bill Dally}.} \bibinfo{year}{2018}\natexlab{}.
\newblock \showarticletitle{Deep Gradient Compression: Reducing the
  Communication Bandwidth for Distributed Training}
  \emph{(\bibinfo{series}{ICLR'18})}.
\newblock
\urldef\tempurl%
\url{https://openreview.net/forum?id=SkhQHMW0W}
\showURL{%
\tempurl}


\bibitem[Liu et~al\mbox{.}(2021)]%
        {liu2021fate}
\bibfield{author}{\bibinfo{person}{Yang Liu}, \bibinfo{person}{Tao Fan},
  \bibinfo{person}{Tianjian Chen}, \bibinfo{person}{Qian Xu}, {and}
  \bibinfo{person}{Qiang Yang}.} \bibinfo{year}{2021}\natexlab{}.
\newblock \showarticletitle{FATE: An Industrial Grade Platform for
  Collaborative Learning With Data Protection.}
\newblock \bibinfo{journal}{\emph{J. Mach. Learn. Res.}} \bibinfo{volume}{22},
  \bibinfo{number}{226} (\bibinfo{year}{2021}).
\newblock
\urldef\tempurl%
\url{http://jmlr.org/papers/v22/20-815.html}
\showURL{%
\tempurl}


\bibitem[McMahan et~al\mbox{.}(2017)]%
        {mcmahan2017communication}
\bibfield{author}{\bibinfo{person}{Brendan McMahan}, \bibinfo{person}{Eider
  Moore}, \bibinfo{person}{Daniel Ramage}, \bibinfo{person}{Seth Hampson},
  {and} \bibinfo{person}{Blaise~Aguera y Arcas}.}
  \bibinfo{year}{2017}\natexlab{}.
\newblock \showarticletitle{Communication-efficient learning of deep networks
  from decentralized data} \emph{(\bibinfo{series}{AISTATS'17})}.
\newblock
\urldef\tempurl%
\url{https://proceedings.mlr.press/v54/mcmahan17a/mcmahan17a.pdf}
\showURL{%
\tempurl}


\bibitem[{Message Passing Interface Forum}(2021)]%
        {mpi40}
\bibfield{author}{\bibinfo{person}{{Message Passing Interface Forum}}.}
  \bibinfo{year}{2021}\natexlab{}.
\newblock \bibinfo{booktitle}{\emph{{MPI}: A Message-Passing Interface Standard
  Version 4.0}}.
\newblock
\urldef\tempurl%
\url{https://www.mpi-forum.org/docs/mpi-4.0/mpi40-report.pdf}
\showURL{%
\tempurl}


\bibitem[Pappas et~al\mbox{.}(2021)]%
        {pappas:2021:ipls}
\bibfield{author}{\bibinfo{person}{Christodoulos Pappas},
  \bibinfo{person}{Dimitris Chatzopoulos}, \bibinfo{person}{Spyros Lalis},
  {and} \bibinfo{person}{Manolis Vavalis}.} \bibinfo{year}{2021}\natexlab{}.
\newblock \showarticletitle{IPLS: A Framework for Decentralized Federated
  Learning} \emph{(\bibinfo{series}{IFIP Networking'21})}.
\newblock
\urldef\tempurl%
\url{https://doi.org/10.23919/IFIPNetworking52078.2021.9472790}
\showDOI{\tempurl}


\bibitem[Paszke et~al\mbox{.}(2019)]%
        {torchSoftware}
\bibfield{author}{\bibinfo{person}{Adam Paszke}, \bibinfo{person}{Sam Gross},
  \bibinfo{person}{Francisco Massa}, \bibinfo{person}{Adam Lerer},
  \bibinfo{person}{James Bradbury}, \bibinfo{person}{Gregory Chanan},
  \bibinfo{person}{Trevor Killeen}, \bibinfo{person}{Zeming Lin},
  \bibinfo{person}{Natalia Gimelshein}, \bibinfo{person}{Luca Antiga},
  \bibinfo{person}{Alban Desmaison}, \bibinfo{person}{Andreas Kopf},
  \bibinfo{person}{Edward Yang}, \bibinfo{person}{Zachary DeVito},
  \bibinfo{person}{Martin Raison}, \bibinfo{person}{Alykhan Tejani},
  \bibinfo{person}{Sasank Chilamkurthy}, \bibinfo{person}{Benoit Steiner},
  \bibinfo{person}{Lu Fang}, \bibinfo{person}{Junjie Bai}, {and}
  \bibinfo{person}{Soumith Chintala}.} \bibinfo{year}{2019}\natexlab{}.
\newblock \showarticletitle{PyTorch: An Imperative Style, High-Performance Deep
  Learning Library}.
\newblock In \bibinfo{booktitle}{\emph{NeurIPS'19}}.
\newblock
\urldef\tempurl%
\url{https://proceedings.neurips.cc/paper_files/paper/2019/file/bdbca288fee7f92f2bfa9f7012727740-Paper.pdf}
\showURL{%
\tempurl}


\bibitem[Roth et~al\mbox{.}(2022)]%
        {roth2022nvidia}
\bibfield{author}{\bibinfo{person}{Holger~R Roth}, \bibinfo{person}{Yan Cheng},
  \bibinfo{person}{Yuhong Wen}, \bibinfo{person}{Isaac Yang},
  \bibinfo{person}{Ziyue Xu}, \bibinfo{person}{YuanTing Hsieh},
  \bibinfo{person}{Kristopher Kersten}, \bibinfo{person}{Ahmed Harouni},
  \bibinfo{person}{Can Zhao}, \bibinfo{person}{Kevin Lu},
  \bibinfo{person}{Zhihong Zhang}, \bibinfo{person}{Wenqi Li},
  \bibinfo{person}{Andriy Myronenko}, \bibinfo{person}{Dong Yang},
  \bibinfo{person}{Sean Yang}, \bibinfo{person}{Nicola Rieke},
  \bibinfo{person}{Abood Quraini}, \bibinfo{person}{Chester Chen},
  \bibinfo{person}{Daguang Xu}, \bibinfo{person}{Nic Ma},
  \bibinfo{person}{Prerna Dogra}, \bibinfo{person}{Mona~G Flores}, {and}
  \bibinfo{person}{Andrew Feng}.} \bibinfo{year}{2022}\natexlab{}.
\newblock \showarticletitle{{NVIDIA} {FLARE}: Federated Learning from
  Simulation to Real-World}. In \bibinfo{booktitle}{\emph{Workshop on Federated
  Learning: Recent Advances and New Challenges}}.
\newblock
\urldef\tempurl%
\url{https://openreview.net/forum?id=hD9QaIQTL_f}
\showURL{%
\tempurl}


\bibitem[Sharma et~al\mbox{.}(2022)]%
        {sacs:2022:decpy}
\bibfield{author}{\bibinfo{person}{Rishi Sharma} {et~al\mbox{.}}}
  \bibinfo{year}{2022}\natexlab{}.
\newblock \bibinfo{title}{decentralizepy: An open-source decentralized learning
  research framework}.
\newblock
\newblock
\urldef\tempurl%
\url{https://github.com/sacs-epfl/decentralizepy}
\showURL{%
\tempurl}


\bibitem[Strom(2015)]%
        {strom2015scalable}
\bibfield{author}{\bibinfo{person}{Nikko Strom}.}
  \bibinfo{year}{2015}\natexlab{}.
\newblock \showarticletitle{Scalable distributed DNN training using commodity
  GPU cloud computing}. In \bibinfo{booktitle}{\emph{16th Annual Conference of
  the International Speech Communication Association}}
  \emph{(\bibinfo{series}{INTERSPEECH'15})}.
\newblock
\urldef\tempurl%
\url{https://www.isca-speech.org/archive_v0/interspeech_2015/papers/i15_1488.pdf}
\showURL{%
\tempurl}


\bibitem[Vogels et~al\mbox{.}(2022)]%
        {vogels2022beyond}
\bibfield{author}{\bibinfo{person}{Thijs Vogels}, \bibinfo{person}{Hadrien
  Hendrikx}, {and} \bibinfo{person}{Martin Jaggi}.}
  \bibinfo{year}{2022}\natexlab{}.
\newblock \showarticletitle{Beyond spectral gap: the role of the topology in
  decentralized learning} \emph{(\bibinfo{series}{NeurIPS'22})}.
\newblock
\urldef\tempurl%
\url{https://proceedings.neurips.cc/paper_files/paper/2022/file/61162d94822d468ee6e92803340f2040-Paper-Conference.pdf}
\showURL{%
\tempurl}


\bibitem[Vogels et~al\mbox{.}(2020)]%
        {vogels2020powergossip}
\bibfield{author}{\bibinfo{person}{Thijs Vogels}, \bibinfo{person}{Sai~Praneeth
  Karimireddy}, {and} \bibinfo{person}{Martin Jaggi}.}
  \bibinfo{year}{2020}\natexlab{}.
\newblock \showarticletitle{Practical Low-Rank Communication Compression in
  Decentralized Deep Learning} \emph{(\bibinfo{series}{NeurIPS'20})}.
\newblock
\urldef\tempurl%
\url{https://proceedings.neurips.cc/paper_files/paper/2020/file/a376802c0811f1b9088828288eb0d3f0-Paper.pdf}
\showURL{%
\tempurl}


\bibitem[Vujasinovic(2023)]%
        {vujasinovic2023secure}
\bibfield{author}{\bibinfo{person}{Milos Vujasinovic}.}
  \bibinfo{year}{2023}\natexlab{}.
\newblock \emph{\bibinfo{title}{Secure Aggregation on Sparse Models in
  Decentralized Learning Systems}}.
\newblock \bibinfo{thesistype}{Master's\ thesis}. \bibinfo{school}{EPFL}.
\newblock
\urldef\tempurl%
\url{https://www.epfl.ch/labs/sacs/wp-content/uploads/2023/02/Secure_Aggregation_on_Sparse_Models_in_Decentralized_Learning_Systems___Milos_Vujasinovic.pdf}
\showURL{%
\tempurl}


\bibitem[Xiao et~al\mbox{.}(2007)]%
        {metropolis}
\bibfield{author}{\bibinfo{person}{Lin Xiao}, \bibinfo{person}{Stephen Boyd},
  {and} \bibinfo{person}{Seung-Jean Kim}.} \bibinfo{year}{2007}\natexlab{}.
\newblock \showarticletitle{Distributed average consensus with
  least-mean-square deviation}.
\newblock \bibinfo{journal}{\emph{J. Parallel and Distrib. Comput.}}
  \bibinfo{volume}{67}, \bibinfo{number}{1} (\bibinfo{year}{2007}).
\newblock
\urldef\tempurl%
\url{https://doi.org/10.1016/j.jpdc.2006.08.010}
\showDOI{\tempurl}


\bibitem[Yang et~al\mbox{.}(2018)]%
        {yang2018applied}
\bibfield{author}{\bibinfo{person}{Timothy Yang}, \bibinfo{person}{Galen
  Andrew}, \bibinfo{person}{Hubert Eichner}, \bibinfo{person}{Haicheng Sun},
  \bibinfo{person}{Wei Li}, \bibinfo{person}{Nicholas Kong},
  \bibinfo{person}{Daniel Ramage}, {and} \bibinfo{person}{Fran{\c{c}}oise
  Beaufays}.} \bibinfo{year}{2018}\natexlab{}.
\newblock \showarticletitle{Applied federated learning: Improving google
  keyboard query suggestions}.
\newblock  (\bibinfo{year}{2018}).
\newblock
\showeprint{1812.02903}


\bibitem[Zhu et~al\mbox{.}(2022)]%
        {zhu2022topology}
\bibfield{author}{\bibinfo{person}{Tongtian Zhu}, \bibinfo{person}{Fengxiang
  He}, \bibinfo{person}{Lan Zhang}, \bibinfo{person}{Zhengyang Niu},
  \bibinfo{person}{Mingli Song}, {and} \bibinfo{person}{Dacheng Tao}.}
  \bibinfo{year}{2022}\natexlab{}.
\newblock \showarticletitle{Topology-aware generalization of decentralized
  {SGD}} \emph{(\bibinfo{series}{ICML'22})}.
\newblock
\urldef\tempurl%
\url{https://proceedings.mlr.press/v162/zhu22d.html}
\showURL{%
\tempurl}


\bibitem[Ziller et~al\mbox{.}(2021)]%
        {ziller2021pysyft}
\bibfield{author}{\bibinfo{person}{Alexander Ziller}, \bibinfo{person}{Andrew
  Trask}, \bibinfo{person}{Antonio Lopardo}, {et~al\mbox{.}}}
  \bibinfo{year}{2021}\natexlab{}.
\newblock \showarticletitle{{PySyft}: A library for easy federated learning}.
\newblock In \bibinfo{booktitle}{\emph{Federated Learning Systems}}.
\newblock
\urldef\tempurl%
\url{https://doi.org/10.1007/978-3-030-70604-3_5}
\showDOI{\tempurl}


\end{thebibliography}
\end{document}